\documentclass[12pt]{article}
\usepackage[margin=2cm]{geometry}
\usepackage{comment}
\usepackage{amsmath,amssymb,extarrows,mathtools,graphicx,subfigure,setspace}
\usepackage{cite}
\usepackage{slashed}
\usepackage{color}
\makeatother

\newcommand{\bei}{\begin{itemize}}
	\newcommand{\eni}{\end{itemize}}
\usepackage{fullpage}
\usepackage{amsmath}
\usepackage{amssymb}
\usepackage{graphicx}
\usepackage{wrapfig}
\usepackage{boxedminipage}
\usepackage{epsfig}
\usepackage{setspace}
\usepackage{subfigure}
\usepackage{float}
\usepackage{hyperref}
\usepackage{soul}
\usepackage{enumerate}

\usepackage{cite}
\newcommand{\be}{\begin{equation}}
\newcommand{\bea}{\begin{eqnarray}}
\newcommand{\eea}{\end{eqnarray}}
\newcommand{\ba}{\begin{align}}
\newcommand{\ea}{\end{align}}
\newcommand{\ee}{\end{equation}}

\def\b{\begin{equation}} \def\e{\end{equation}}
\def\bd{\begin{displaystyle}} \def\ed{\end{displaystyle}}
\def\ba{\begin{array}} \def\ea{\end{array}}

\def\bee{\begin{enumerate}}
	\def\eee{\end{enumerate}}

\def\bes{\begin{eqnarray*}}
	\def\ees{\end{eqnarray*}}
\def\be{\begin{eqnarray}}
\def\ee{\end{eqnarray}}

\numberwithin{equation}{section}
\begin{document}
	\onehalfspacing
	\vfill
	\begin{center}
		
		{\Large { Ultimate Limits to Computation: Anharmonic Oscillator} }
		\vspace*{.9cm}

		{Fatemeh Khorasani$^{1}$,
			Mohammad Reza Tanhayi$^{1,\star}$ and Reza Pirmoradian$^{2}
		$}
		\vspace*{.5cm}
		
		{$^{1}$Department of physics, Central Tehran Branch, Islamic Azad University, Tehran, Iran\\
	$^{2}$ School of Particles and Accelerators, Institute for Research in Fundamental Sciences (IPM)
	P.O. Box 19395-5531, Tehran, Iran	}
		
		\vspace*{0.5cm}
		{E-mail: {\tt  mtanhayi@ipm.ir  }}%
		\vspace*{1cm}
	\end{center}

	\begin{abstract}
		
		Motivated by studies of ultimate speed of computers, we examine the question of minimum time of orthogonalization in a simple anharmonic oscillator and  find an upper bound on the rate of computations. Furthermore, we investigate the growth rate of complexity of operation when the system undergoes a definite perturbation. At the phase space of the parameters, by numerical analysis, we find the critical point where beyond that the rate of complexity changes its behavior.

	\end{abstract}
	\textbf{keywords:} Complexity; Orthogonality; Anharmonic Oscillator \\ 
	\vspace{0.5cm} 
\section{Introduction}
 How fast can a computer be? Is there any limitation to the data storage? These are some interesting and fundamental questions  that yet need to be addressed in the relevant contexts of theoretical physics and computer science. In fact, a number of physical and practical factors involves in the limitation of computation,  as well as, the data storage. There are some approaches employ the laws of physics to investigate these issues, for example, the energy and the number of degrees of freedom that the system could achieve, can in principle control the speed limit and the amount of memory limitations, respectively.   \\ 
 To be specific, let us consider  speed limit implemented by quantum mechanics. The speed limit might be defined by the minimal time a system needs to evolve from an initial state to a final orthogonal state \cite{fast:quantum}.  For a quantum system with energy $E$, there is a limitation in the quantum computation which is identified by  the minimum time needed for a given state to evolve to an orthogonal state.  As a matter of fact, the famous time-energy uncertainty relation and its Margolus-Levitin interpretation of the average energy  \cite{aaaa}, can limit the speed of evolution of a system. This theorem imposes a fundamental upper limit on the computational speed for a computer. A simple step of computation done by a computer can be translated by passing one state to its orthogonal state via successive application of logic gates. Any performing of a task or equivalently  passing through a gate needs $\Delta t$ time which is controlled by $\Delta t\ge \frac{\pi\hbar}{2\Delta E}$. In this sense, one can argue that quantum mechanics imposes a limit on the speed of a dynamical evolution.  This limitation on the speed of processing can be translated into the maximum number of distinct states that a given system passes per unit of time. If $E$ be the instantaneous energy of the system  throughout the computation then one can say that an upper
 bound on the number of operations per unit time is given by $$\frac{1}{\Delta t}< \frac{2}{\pi\hbar} E,$$ this is the Lloyd's bound \cite{2000Natur.406.1047L}. This is a limitation for information processing, namely, for a quantum system with a given average energy $E$, the upper limit on the number of operations is given by $\frac{2E}{\pi\hbar}$. The states are orthogonal and in this sense, one can say that an upper bound on the speed of processing is enforced by the quantum mechanics.  This bound has been investigated for quantum computer, and in this way the key point is no need to work with orthogonal states, instead the quantum superpositions of states are used. In this paper, we apply the method of quantum mechanics to a specific problem of calculating the orthogonalization time for a simple anharmonic oscillator. 
 
 On the other hand, in computer science,  the difficulty of doing a physical task has a measure which is named as the complexity. The complexity quantifies the difficulty  of evolving a reference quantum state into another (target) state. The computational complexity of an operation is defined by the minimum number of logical gates that are needed for solving a  problem \cite {wert,C.moor1}. It would be an interesting to study whether the proposed bound has any thing to do with complexity or not, this point has been considered in various papers \cite {Susskind:2014rva,Doroudiani:2019llj,Brown:2015bva,wer,Jefferson:2017sdb,ref1,Chapman:2021jbh}. In Ref. \cite{REZA:tanhayi}, we have studied the effect of both a magnetic and an electric field and investigated the Lloyd's bound. Our main goal in this article is to examine this bound for a quantum system of anharmonic oscillator. 
 
 The layout of this paper is as follows. In Section 2, we calculate the orthogonalization time for a simple anharmonic oscillator and in Section 3, we explore Lloyd's bound for this system. The concluding remarks are given in section 4. Finally, with two appendices we present some mathematical calculations.


\section{Orthogonality Time}

In the field of information processing and quantum measurements, the evolution of a quantum state and also how fast the state can evolve, are two quantitative questions. The orthogonal states can be distinguished from each other, therefore,  a transition from a state to an orthogonal state, might be considered as an elementary step of a computational process. The maximal speed of evolution can be identified by the minimal time that a system needs to evolve from an initial state to another orthogonal state. This time is known as the quantum speed limit time and for a system of total energy $E$, this time is given by the Margolus-Levitin formula \cite{aaaa}. In other words, for a given quantum state, say $\mid\Psi\rangle$, the orthogonality time can be obtained via the following relation 
\bea \label{STau}
S(\tau_{\bot})\equiv \langle\Psi\mid\Psi_{\tau_{\bot}}\rangle=\displaystyle\sum_{n=0}^{\infty}{\mid c_{n}\mid}^2 e^{-iE_{n}{\tau_{\bot}}}=0,\label{st}
\eea
where  $\mid\Psi_{\tau_\bot}\rangle$ is the corresponding orthogonal state, this yields \cite{REZA:tanhayi} $$
\tau_{\bot}=\frac{\pi}{2 \mathbf{E}}.$$ 
The Margolus-Levitin formula gives a fundamental limit on quantum computation, we will come back to this point later, and now, we apply this formula  for a simple harmonic oscillator. Actually what we shall find is the orthogonality time   which leads to the speed of the dynamical evolution. Let us consider the following Hamiltonian in one dimension\footnote{Note that from now on we have used the natural units $\hbar=c=1$.}   
\bea
H=\frac{1}{2m}p^2+\frac{1}{2}m{\omega}^2 x^2+q\mathcal{E}x+\lambda x^4,
\eea
where the effect of a weak external electric field is given by $H_p=q\mathcal{E}x+\lambda x^4$ and $m$ and $q$ stand for the mass and charge of system, respectively. The lowest non-vanishing order of the energy shift is then given by
\bea
E_n= \omega(n+\frac{1}{2})+\frac{3\lambda}{4 m^2 \omega^2}[2 n^2+2n+1]-2m\frac{ \omega_e^2}{\omega^2}
\eea
where $n$ is a positive integer or zero and $\lambda$ is a small positive real parameter, also $\omega_e$ is defined by 
\bea
\frac{q\mathcal{E}}{2m}\equiv \omega_e.
\eea
\begin{figure}[h]\label{Fig1}
	\centering
	\includegraphics[scale=.28
	]{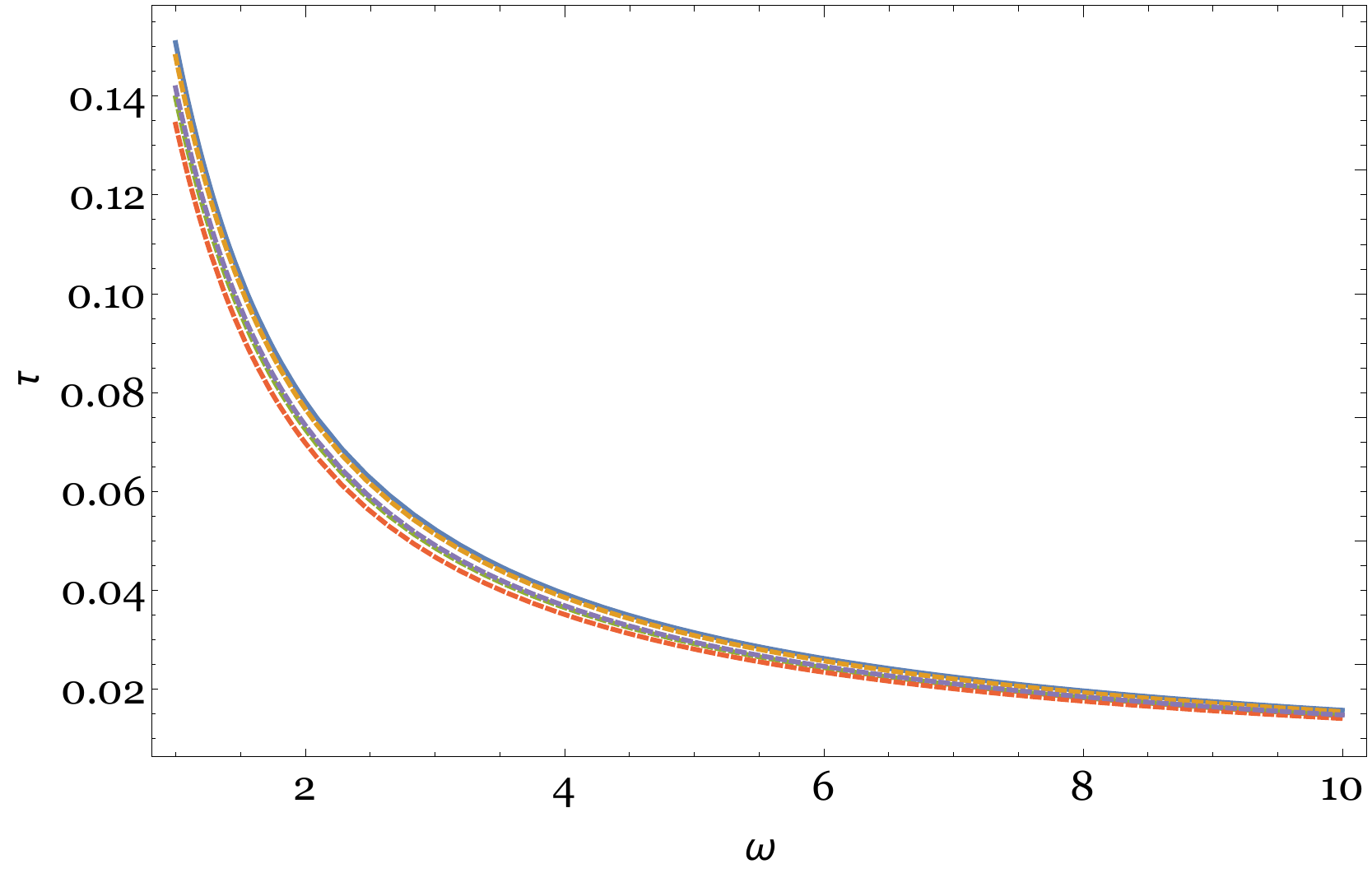}\includegraphics[scale=.28]{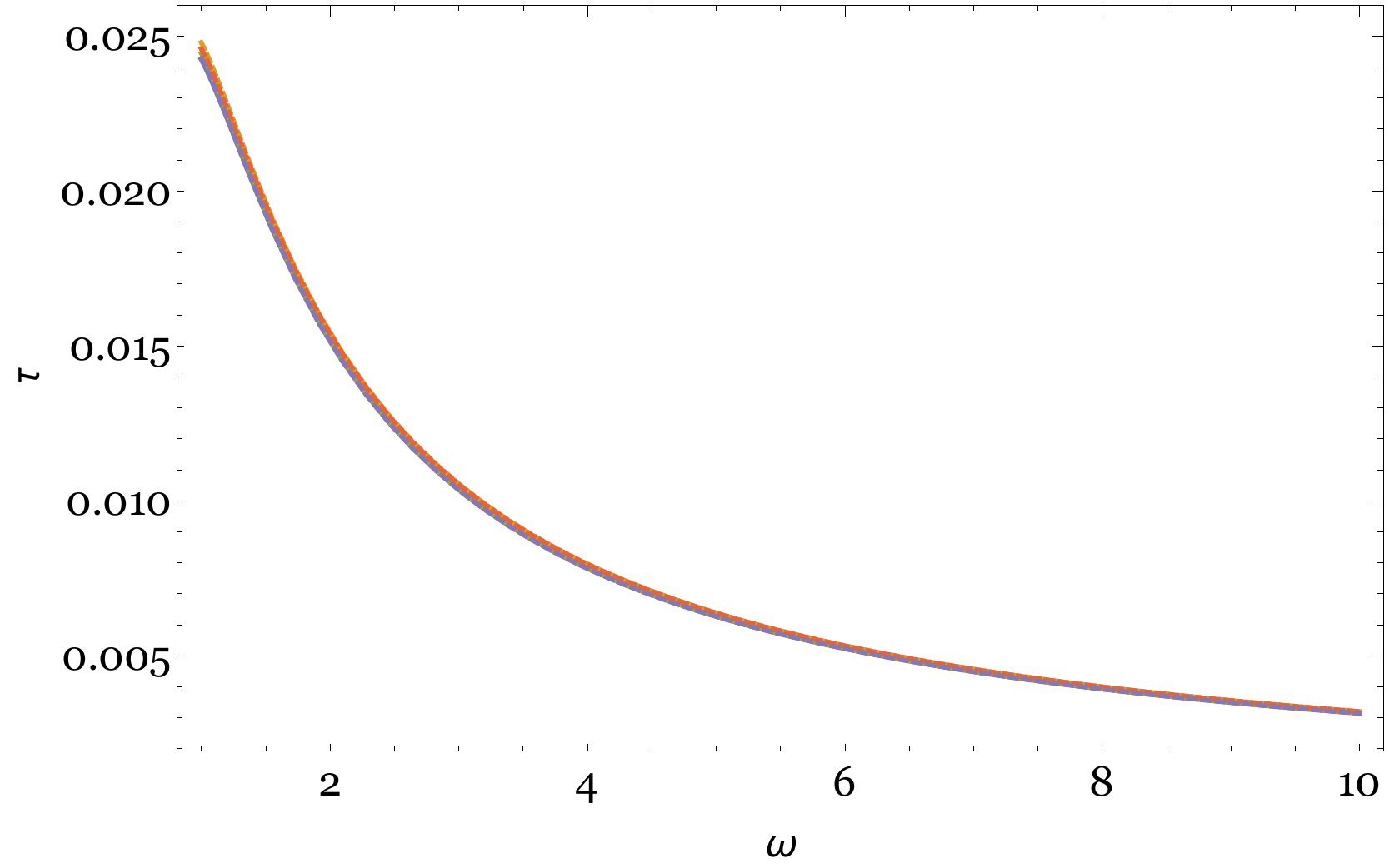}
	\caption{ The orthogonality time for generic coefficient (dashed line) and the case that we have
		chosen in this paper (solid line), for N =20 (left panel) and N = 100 (right	panel). In both figures we set m = 1, $\lambda=3\times10^{-3}$ and $\omega_e=3\times10^{-1}$.}
		 
\end{figure} 
In order to compute the average energy in the state $\mid\Psi\rangle$, let us suppose that our system passes through $N$ mutually orthogonal states in time $\tau$, consequently, the eigenfunction is displayed as follows
\bea\label{wer}
\mid\Psi\rangle=\displaystyle\sum_{n=0}^{N-1}c_n\mid n\rangle.
\eea
For a quartic anharmonic oscillator, the expectation value of energy is given by 
\bea
\mathbf{E}&=&\langle\Psi\mid H\mid\Psi\rangle \cr \nonumber\\
&=&\displaystyle\sum_{n=0}^{N-1}{\mid c_{n}\mid}^2\Big[\omega[n+\frac{1}{2}]+\frac{3\lambda}{4 m^2 \omega^2}[2 n^2+2n+1]-2m\frac{ \omega_e^2}{\omega^2}\Big].
\eea
To make calculations easier, one can start by
 considering the set of evolutions that pass through an exact cycle of $N$ which are supposed to be mutually orthogonal states at a constant rate  in time $\tau$; In this case and for large $N$, one has $ c_{n}=\sqrt{\frac{1}{N}}$, leading to following expression 
\bea
 \mathbf{E}&=&\frac{1}{N}\bigg[\displaystyle\sum_{n=0}^{N-1}\Big(\omega[n+\frac{1}{2}]+\frac{3\lambda}{4 m^2\omega^2}[2 n^2+2n+1]-2m\frac{ \omega_e^2}{\omega^2}\Big)\bigg]\cr \nonumber\\&=&\frac{N  \omega}{2}+\frac{\lambda}{4 m^2 \omega^2}+\frac{N^2 \lambda}{2 m^2\omega^2}-2m\frac{ \omega_e^2}{\omega^2}
\eea
The state defined in \eqref{wer}, with all coefficients equal, has the property of cycling through orthogonal states for the unperturbed  harmonic oscillator, as shown in the Margolus-Levitin paper, but for the perturbed case, this might work as well. For sufficiently large $N$, in figure (1), we have plotted the resultant orthogonality time for generic four random coefficient with $N = 20$ and $N = 100$. As the figure shows for large $N$, the curves coincide to what we have considered in this paper. Moreover, making use the relation given in appendix and also  $ \omega_\lambda^2\equiv\frac{\mathcal{\lambda}}{2m^3}$, one can show that 
\bea\label{T}
\tau_\bot\geq\frac{\pi}{N \omega +m\frac{\omega_\lambda^2}{ \omega^2}+2 m N^2 \frac{ \omega_\lambda^2}{ \omega^2}-4m\frac{ \omega_e^2}{\omega^2}}\,.
\eea 
It is worth to mention that we are dealing with large $N$, and the following limit is supposed to be hold 
	$$\omega_e<\frac{\sqrt{\lambda(\frac{1}{2}+N^2)+m^2 N \omega^3}}{2m^{\frac{3}{2}}},$$
leading to a non-negative energy. According to our definition of the time which takes for a given quantum system  of energy $\mathbf{E}$ to go from one state to an orthogonal state, the relation \eqref{T} shifts the bound to upper bound of the rate of complexity.

\begin{figure}[h]\label{Fig2}
	\centering
	\includegraphics[scale=.28
	]{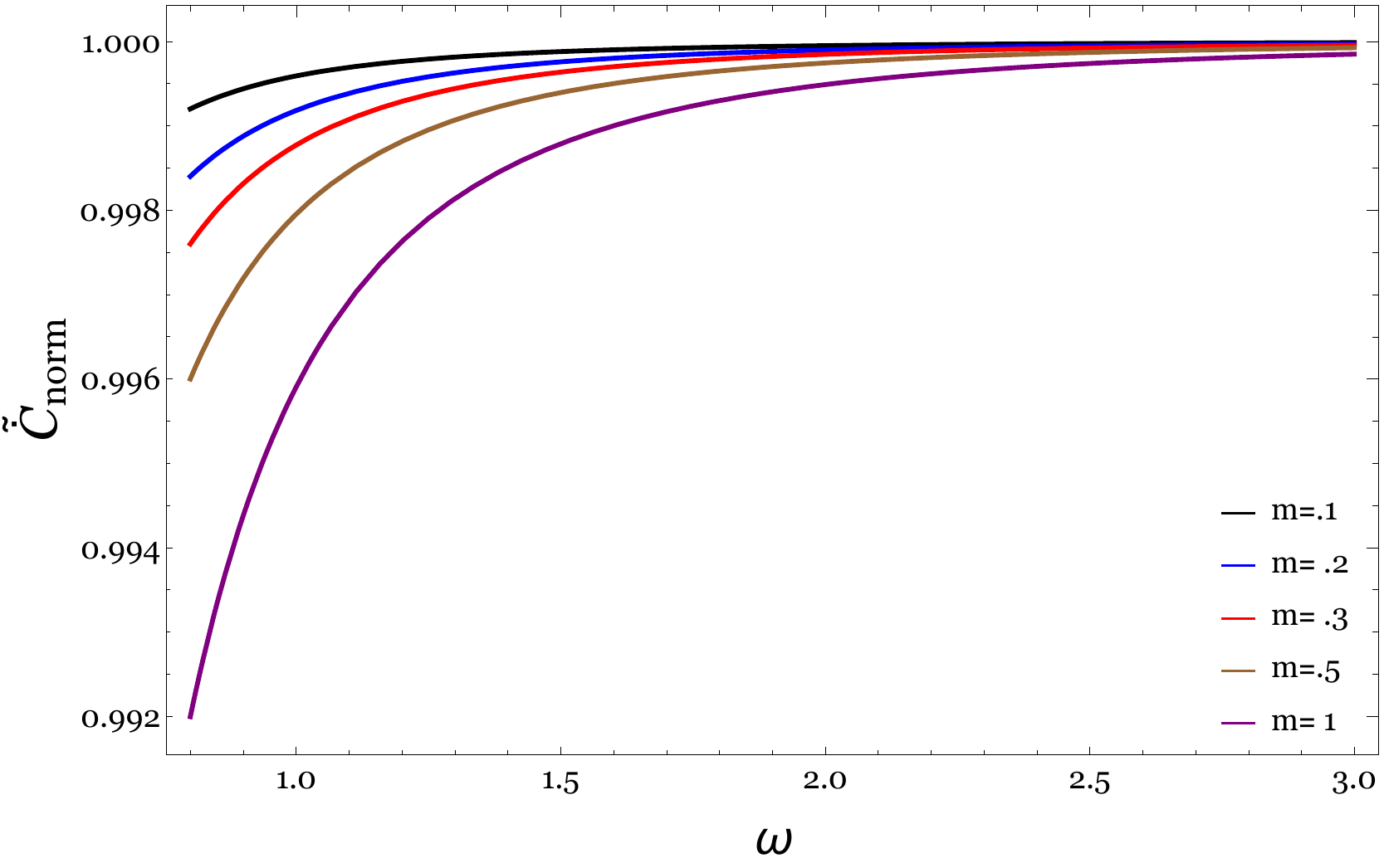}\includegraphics[scale=.28]{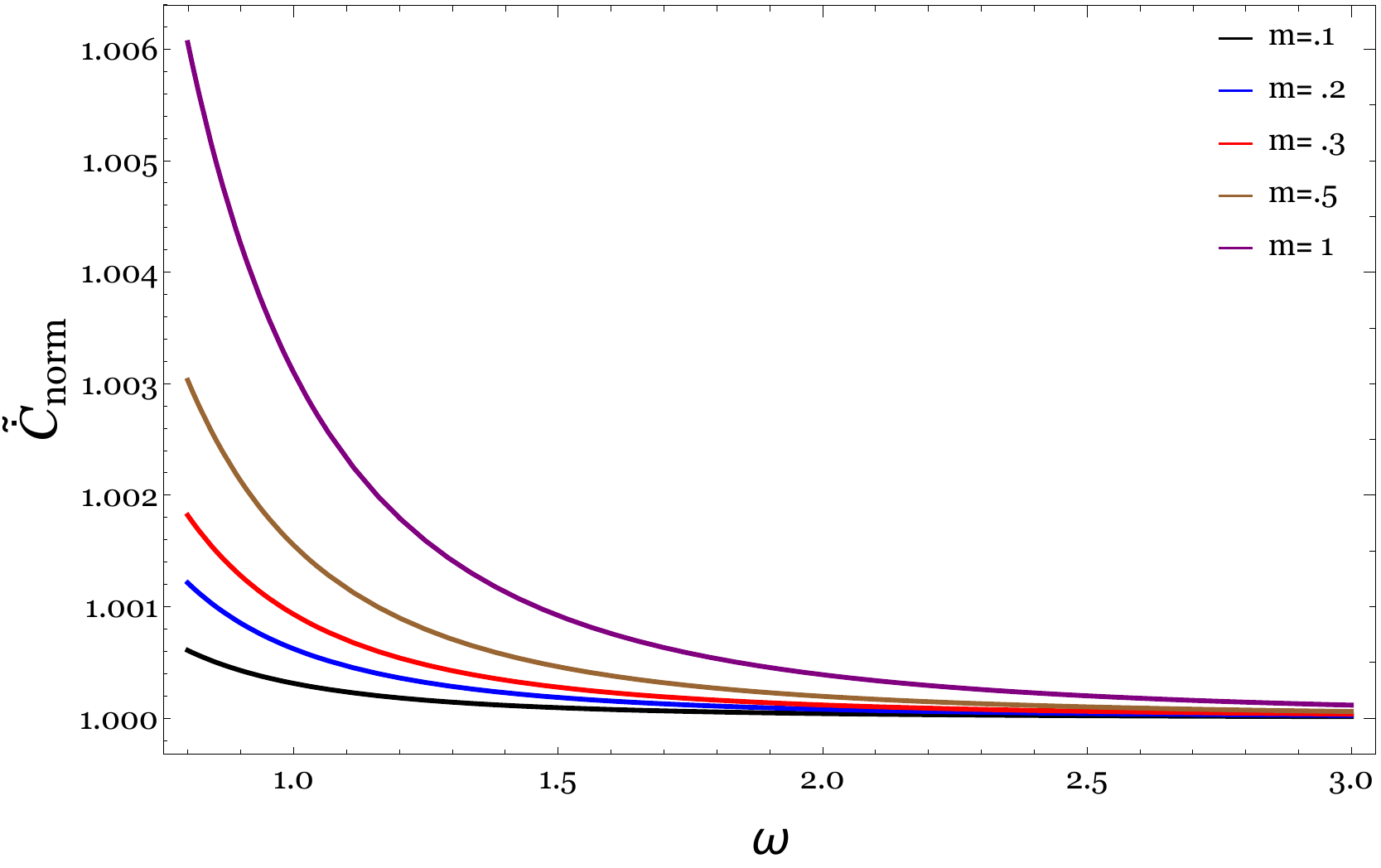}
	\caption{  The rate of normalized complexification, defined by $\tilde{\dot{\mathcal{C}}}_{norm}=\frac{\pi{\dot{\mathcal{C}}}}{N\omega}$, for different values of mass, {\em left panel}: $\omega_\lambda=8\times10^{-3}$, {\em right panel}:  $\omega_\lambda=10^{-2}$.  In both we take $N=100$ and  $\omega_e=65\times10^{-2}$. } 
\end{figure}

\begin{figure}[h]\label{Fig3}
	\centering
	\includegraphics[scale=.35
	]{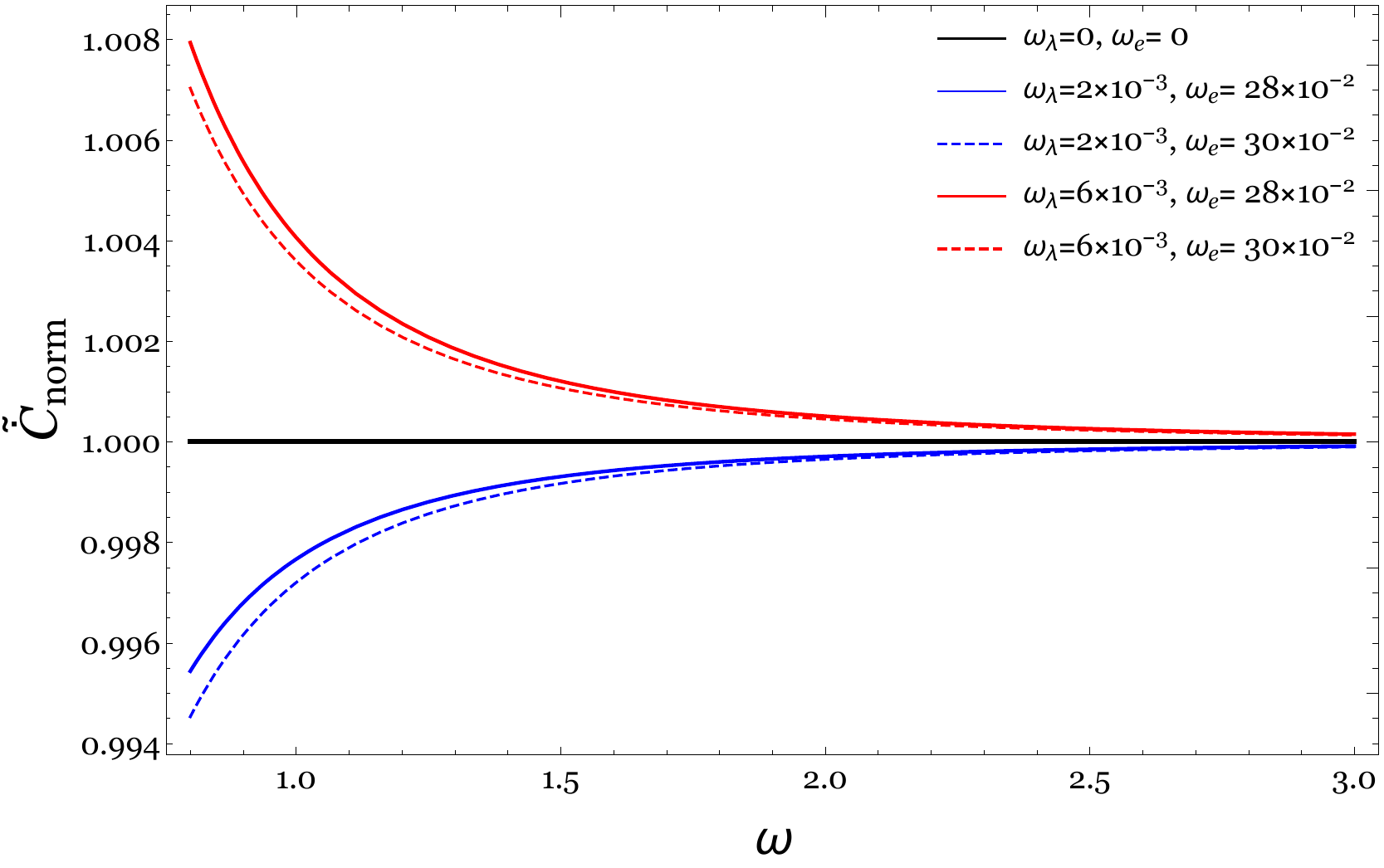}
	\caption{ The rate of normalized complexification, with respect to $\omega_{e}$.  When the anharmonic parameter induction reaches a definite critical value given by Eq.\eqref{criti}, the behavior of this rate changes. Note that we set  $N=100$ , $m=1$. } 
\end{figure}

\section{Upper Bound of Complexification}
 The information process is defined by successive application of logical gates which are needed to transform a given
initial state to a final state. In other words, a logical gate refers to the actual physical device that performs a logical operation and performing any operation takes some time $t$. If one defines a
Hamiltonian action $H_g$ to a task which is done by given gate, then the operation is given by $U(t)=e^{iH_g\,t}$. The unitary evolution takes the initial state $|0\rangle$ to desired final state $U(t)|0\rangle$ after time $t$. Consequently, a sequential application of $n$ gates, are given by \cite{Cottrell:2017ayj}
\begin{equation}
|0\rangle\rightarrow U(t)|0\rangle,\,\,\,\,\,\,\,
U(t)=\mathbf{T} \prod_i U(t_{i+1},t_i), \end{equation}
where $\mathbf{T}$ is the time ordered operator and $U(t_{i+1},t_{i}) $ stand for an orthogonalizing gate. As mentioned, complexity is a measure of the difficulty of doing a physical task and it is defined by the minimum number of logical gates that are needed for solving a  problem \cite {wert,C.moor1}. From the theoretical point of view, complexity is defined by the number of elementary unitary operations which are required to build up a desired final state from a given reference state \cite{Guo:2018kzl}. In the present case, the rate of complexification gets a strong bound given by (see appendix A) 
\begin{equation}\dot{\mathcal{C}}\leq \frac{1}{\tau_{\bot}}\label{cm},\end{equation}
where  $\tau_{\bot}$ is the orthogonalization time of the system. We should mention that the relation (3.2) between the rate of complexification and the orthogonality time is a conjecture, and it works with assuming that the optimal circuit is made of 
orthogonalizing gates\footnote{We thank the referee for his/her comment on this point.}. Therefore, in our case the rate of complexification is given by 
\begin{align}\label{3.3}
\dot{\mathcal{C}}<\frac{1}{\pi}\Big( N \omega +m\frac{\omega_\lambda^2}{ \omega^2}+2 m N^2 \frac{ \omega_\lambda^2}{\omega^2}-4m\frac{\omega_e^2}{\omega^2}\Big)\,
\end{align}
\begin{figure}[h]\label{Fig4}
	\centering
	\includegraphics[scale=.25
	]{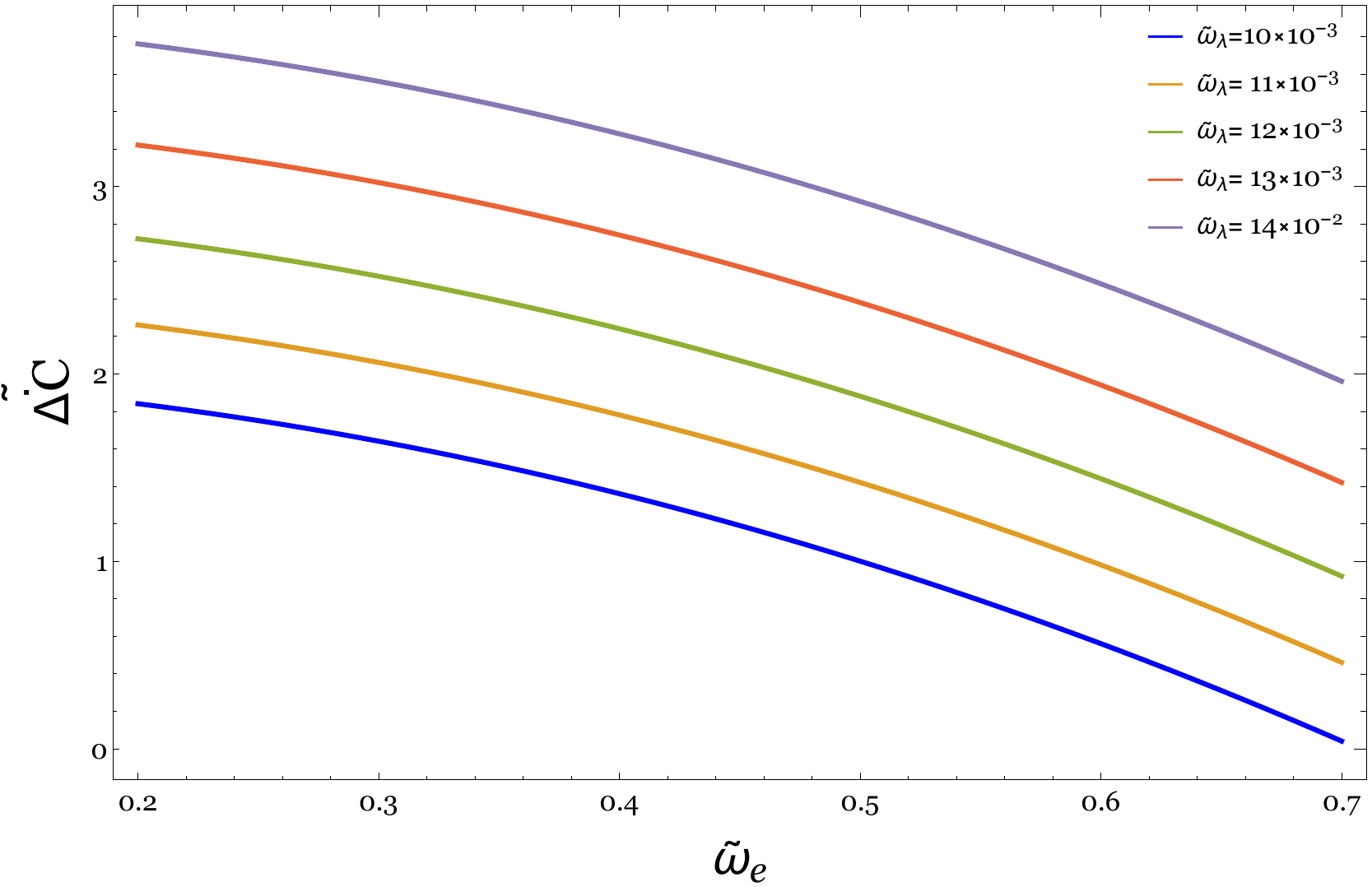}
	\hspace{8mm}
	\includegraphics[scale=.20
	]{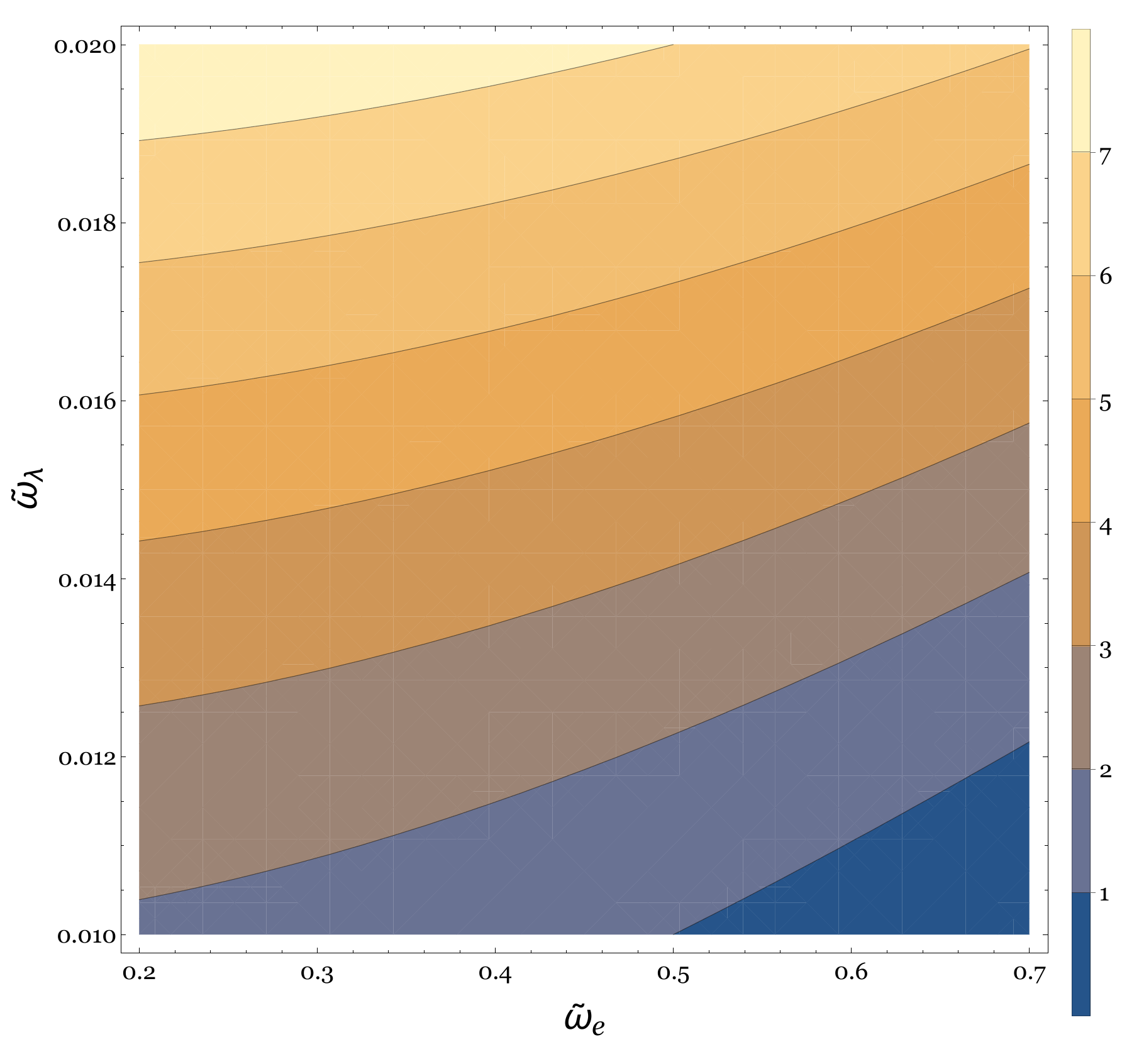}
	\caption{The effect of anharmonic parameter and electric field on the rate of complexity ($\widetilde{\Delta{\dot{\mathcal{C}}}}=\frac{\pi{\dot{\mathcal{C}}}-N \omega}{m}$).  The rate of complexity increases by $\tilde{\omega}_{\lambda}$ while it decreases by increasing  $\tilde{\omega}_{e}$. Note that we set  $N=100$ , $\widetilde{\omega}_\lambda\equiv\frac{\omega_\lambda}{\omega} $ and $\widetilde{\omega}_e\equiv\frac{\omega_e}{\omega}$. }
\end{figure}
In figure (3) we have plotted the rate of complexity for different values of $\omega_\lambda$ and $\omega_e$.  As it is shown from figures (2) and (3), there is a critical value for anharmonic parameter which beyond that the rate of complexity changes its behavior drastically, it is given by
\begin{equation}\omega_\lambda^{cri}=\frac{2\omega_e}{\sqrt{1+2N^2}}\,\,.\label{criti}
\end{equation}
The results indicate that the normalized complexification of the charged anharmonic oscillator saturates to a definite value for large $\omega$. However, there is a critical value for anharmonic parameter  which defines the behavior of this saturation. In figure (4) we show that the electric field can completely neutralize the additive effect of $\lambda$, so that when  Eq.\eqref{criti} is fully satisfied, the complexification does not change, namely one has $\widetilde{\Delta{\dot{\mathcal{C}}}}=0$. It is worth to mention that these results of the complexification rate are based on the assumption that the inequality \eqref{3.3} is saturated.

If we want to make a simple comparison with a charged harmonic oscillator in the presence of electric and magnetic fields, it reveals that for charged anharmonic oscillator, the electric field has a decreasing effect on the rate of complexity. This observation is the same for charged harmonic oscillator. Instead, figure (4) indicates that the  anharmonic parameter, like the magnetic field in the charged harmonic oscillator, plays an effective role in increasing  the rate of complexity.

\section{Remarks and Conclusions}

 Indeed, computation complexity, particularly its quantum version, is a difficult subject, and in order to compute the complexity, one needs to choose a proper reference and target states, then the next important issue is to identify a set of unitary gates with which to construct the desired unitary operator $U$,  which acts on reference state to built target state. In fact, the complexity of the target state is defined as
the minimum number of gates needed to construct such a unitary operator.  There are some proposals to choose an optimal number of gates, say as Nielsen geometric model \cite{Nil} and Fubini-Study approach (see \cite{ Jefferson:2017sdb}). 
In this paper, however, our main focus was somewhat different, we were interested on the quantitative behavior of complexity as a system
evolves, albeit in a particular case, where the operation sends states to the orthogonal  states. Our method based on the Margolus-Levitin and Lloyd proposals. The Margolus-Levitin theorem sets a limit on quantum computation by identifying the minimum time for a quantum system with an energy $E$ to go from one state to an orthogonal state. There is a limitation or an upper bound on the rate of computations which is known as Lloyd's bound and it deals with the minimal time to perform a task done by a physical device. The minimal time is controlled by the energy $E=\langle H \rangle$, meaning that the energy of a system sets a limitation of computation, therefore, the complexity is affect by the energy of the system. The processing of information might be given by  the time rate of change of the complexity. In our previous work, we studied the  complexity of a charged harmonic oscillator in the presence of both magnetic and electric fields and observed that the time of orthogonality for small and large values of magnetic field behaves differently, and also, the rate of complexity increases/decreases by turning on the magnetic/electric field. In this paper, for a simple anharmonic oscillator, we studied the minimal time of orthogonalization and observed that the average value of the energy  plays a crucial role in this process. For the perturbation as $q\mathcal{E}x+\lambda x^4$, we investigated the effect of introduced parameters on the rate of complexification. Numerical calculation shows that massive systems are more complex, however, the system has a critical value of the parameters $\lambda$ and $\mathcal{E}$, where the situation reverses. Beyond the critical value by increasing the mass, the rate of complexity decreases.  We showed that the rate of complexity increases by $\lambda$ while it decreases by increasing   $\mathcal{E}$, however, we found the critical value as follows  
$$\lambda=\frac{2mq^2}{1+2N^2}\mathcal{E}^2$$
where at this point, one observes that the complexification does not change, and this is the point where the system undergoes a phase transition.   Moreover, the numerical computation showed that for those perturbation up to all orders, that have odd power of $x$, the rate of  complexity decreases whereas the even power increases this rate and leads to more complexification.

\section*{Acknowledgment }

R.P would like to kindly thank A. Naseh  and B. Taghavi for useful comments and discussions on related topics. 



\section*{Appendix A}
In this appendix we briefly review the arguments underlying the relation \eqref{cm} and for more details, we refer to \cite{Cottrell:2017ayj}. Let us suppose for a a single step of operation, the gate $G$, takes some time $\Delta t$ to perform its task. Within the quantum mechanical language, one can say that $G$ is implemented by a Hamiltonian action as $G = e^{iH\Delta}$, with an energy $E = \langle H\rangle$, and the Hamiltonian of the gate is given by $H$. The time of performing a task has been controlled by $E$, namely, higher $E$ results less time for the gate to perform its task. The performance time of a task is supposed to be $T \equiv n\Delta t$ and the actual computation time of the system is given by $t$, then one has \begin{equation}t\ge T \label{41}\end{equation}
As a matter of fact, $T$ is the time that it takes for the fastest classical computer to implement a given task.  On the other hand according to the Margolus-Levitin theorem, each gate must take at least the minimum time to perform its task. and for each time step one has \be \Delta t\ge \frac{\pi}{2E}\ee
this can be rearranged  to the following form 
\be \frac{1}{\Delta t}\leq \frac{2E}{\pi}\ee
 an upper
which the above relation impose a bound on the instantaneous number of operations per unit time. The actual complexity of the state must be lower than $n$, the number of gates in the fastest classical computer that produces it, then making use of the assumption \eqref{41}  one
obtains a bound in the rate of complexity change as follows 
\be \dot {\cal C}\equiv \frac{\Delta{\cal C}}{\Delta t}\ge \frac{\pi}{2E}.\ee
It is also worth to mention that our assumption of orthogonal states is based on the Lloyd approach, where in his approach,  the gates are all quantum-mechanical implementations
of classical logic gates acting on ‘classical’ states. Two different classical states are supposed to be orthogonal, then all the gates are chosen to evolve the quantum states they encounter to
orthogonal ones, as well. In other words let us assume that the time evolution is implementing a series computation which is given by $$U(t)=\mathbf{T} \prod_i U(t_{i+1},t_i),$$ 
and each evolution refers to an orthogonalizing gate. Therefore, one can say that the state remains orthogonal when the complexity of the circuit increases by one unit. On the other hand by definition the circuit complexity increases faster than the state complexity which this gives us the relation \eqref{cm}.

\section*{Appendix B: Eigenvalue for perturbation theory }\label{apppa}
 In this appendix, we find the energy change of the simple anharmonic oscillator whose Hamiltonian is given by 
\bea\label{aaa}
{H}=\frac{1}{2m}p^2+\frac{1}{2}m{\omega}^2 x^2+\lambda_o x^{2k-1}+\lambda_e x^{2k},
\eea
where $0\leq\lambda_o,\lambda_e\ll1$ and $k$ is a positive integer number. It is shown that the growth rate of complexity for an anharmonic oscillator for the even order perturbation, increases while for the odd order the rate has a decreasing behavior. The first-order change in the odd perturbation vanishes, on the other hand, the second-order correction to the energy becomes 
\bea
E^2_{n_{odd}}=-\lambda^2_o\sum_{\substack{	m\neq n 		
}}\frac{\mid \langle m \mid  x^{2k-1} \mid n\rangle\mid^2}{E_{mn}}
\eea
which is negative, while one gets a positive number for even perturbation  	\bea
E^1_{n_{ even}}=\lambda_e\langle n\mid  x^{2k} \mid n\rangle,
\eea
where the position operator $x$ is given by
\bea
x=\sqrt{\frac{1}{2m\omega}}(a+{a}^{\dagger}).
\eea
Not let us calculate the first-order corrections for a
harmonic oscillator with applied perturbation say as $\lambda_1 x,\lambda_2 x^2,\lambda_3 x^3,...$, which leads to
\bea
\langle m \mid (a+{a}^{\dagger})
 \mid n\rangle\mid&=&\sqrt{n} \delta _{m,n-1}+\sqrt{n+1} \delta _{m,n+1}
\cr \nonumber\\
\langle m \mid (a+{a}^{\dagger})^2
\mid n\rangle\mid&=&2 n \delta _{m,n}+\delta _{m,n}+\sqrt{n-1} \sqrt{n} \delta _{m,n-2}+\sqrt{n+1} \sqrt{n+2} \delta _{m,n+2}
\cr \nonumber\\
\langle m \mid (a+{a}^{\dagger})^3
\mid n\rangle\mid&=&3 n^{3/2} \delta _{m,n-1}+3 \sqrt{n+1} n \delta _{m,n+1}+\sqrt{n-2} \sqrt{n-1} \sqrt{n} \delta _{m,n-3}
\cr \nonumber\\
&\,\,\,\,\,\,\,\,\,+&3 \sqrt{n+1} \delta _{m,n+1}+\sqrt{n+1} \sqrt{n+2} \sqrt{n+3} \delta _{m,n+3}
\eea
After doing some algebra one also finds 
\bea
\langle n\mid  x^{2} \mid n\rangle&=&\frac{2 n+1}{2 m\omega}\cr \nonumber\\
\langle n\mid  x^{4} \mid n\rangle&=&\frac{3 \left(2 n^2+2 n+1\right)}{4 m^2\omega^2}\cr \nonumber\\
\langle n\mid  x^{6} \mid n\rangle&=&\frac{5 (2 n+1) \left(2 n^2+2 n+3\right)}{8 m^3\omega^3}
\cr \nonumber\\
\langle n\mid  x^{8} \mid n\rangle&=&\frac{35 \left(2 n^4+4 n^3+10 n^2+8 n+3\right)}{16 m^4\omega^4}
\cr \nonumber\\
\langle n\mid  x^{10} \mid n\rangle&=&\frac{63 (2 n+1) \left(2 n^4+4 n^3+18 n^2+16 n+15\right)}{32 m^5 \omega^5}
\cr \nonumber\\
\eea
We can also get energy corrections up to the second level
\bea\label{Eodd}
\sum_{\substack{	m\neq n 		
}}\frac{\mid \langle m \mid  x \mid n\rangle\mid^2}{E_{mn}}&=&\frac{1}{(2 m\omega)\omega}
\cr \nonumber\\
\sum_{\substack{	m\neq n 		
}}\frac{\mid \langle m \mid  x^3 \mid n\rangle\mid^2}{E_{mn}}&=&\frac{30 n^2+30 n+11}{(2 m\omega)^3 \omega}
\cr \nonumber\\
\sum_{\substack{	m\neq n 		
}}\frac{\mid \langle m \mid  x^5 \mid n\rangle\mid^2}{E_{mn}}&=&\frac{630 n^4+1260 n^3+2030 n^2+1400 n+449}{(2 m\omega)^5 \omega}
\cr \nonumber\\
\sum_{\substack{	m\neq n 		
}}\frac{\mid \langle m \mid  x^7 \mid n\rangle\mid^2}{E_{mn}}&=&\frac{3 \left(4004 n^6+12012 n^5+42350 n^4+64680 n^3+81788 n^2+51450 n+14793\right)}{(2 m\omega)^7 \omega}
\cr \nonumber\\
\eea
These calculations indicate that the rate of complexity increases (decreases)  by the even (odd) order of perturbation. 
\section*{Data Availability Statement}
All data generated or analysed during this study are included in this published article (and its supplementary information files).

\end{document}